# A Critical Review of Safe Reinforcement Learning Techniques in Smart Grid Applications

Van-Hai Bui[1,*], Srijita Das[2], Akhtar Hussain[3], Guilherme Vieira Hollweg[1], and Wencong Su[1,*]

[1]Department of Electrical and Computer Engineering, University of Michigan-Dearborn, USA.
[2]Department of Computer and Information Science, University of Michigan-Dearborn, USA.
[3]Department of Electrical and Computer Engineering, Laval University, Canada.

*Abstract*—The high penetration of distributed energy resources (DERs) in modern smart power systems introduces unforeseen uncertainties for the electricity sector, leading to increased complexity and difficulty in the operation and control of power systems. As a cutting-edge machine learning technology, deep reinforcement learning (DRL) has been widely implemented in recent years to handle the uncertainty in power systems. However, in critical infrastructures such as power systems, safety issues always receive top priority, while DRL may not always meet the safety requirements of power system operators. The concept of safe reinforcement learning (safe RL) is emerging as a potential solution to overcome the shortcomings of conventional DRL in the operation and control of power systems. This study provides a rigorous review of the latest research efforts focused on safe RL to derive power system control policies while accounting for the unique safety requirements of power grids. Furthermore, this study highlights various safe RL algorithms applied in diverse applications within the power system sector, from single grid-connected power converters, residential smart homes, and buildings to large power distribution networks. For all methods outlined, a discussion on their bottlenecks, research challenges, and potential opportunities in the operation and control of power system applications is also presented. This review aims to support research in the area of safe RL algorithms, embracing smart power system operation with safety constraints amid high uncertainty from DERs.

## ABBREVIATIONS

| | |
|---|---|
| AC | Alternating current |
| CIG | Converter-interfaced generators |
| CMDP | Constrained Markov decision process |
| DC | Direct current |
| DER | Distributed energy resource |
| DNN | Deep neural network |
| DRL | Deep reinforcement learning |
| EV | Electric vehicle |
| ESS | Energy storage system |
| HVAC | Heating, ventilation, and air conditioning |
| MDP | Markov decision process |
| MG | Microgrid |
| ML | Machine learning |
| LLM | Large language models |
| RES | Renewable energy sources |
| VVC | Volt-VAR control |
| V2G | Vehicle-to-grid |

## I. INTRODUCTION

THE power and energy systems have undergone significant transformations in recent years, transitioning from conventional setups to modern, dynamic systems characterized by high penetration of distributed energy resources (DERs) [1], [2]. Conventional power systems relied heavily on centralized, model-based approaches for the system operation and control. For instance, various optimization problems in the operation of power systems could be solved using mixed-integer linear programming [3], [4] or quadratic programming [5] for deterministic environments. To account for uncertainty factors, robust optimization-based models [6] and stochastic optimization-based models [7], [8] have been widely applied in the literature. Additionally, model predictive control-based optimization models [9] have been developed for power system modeling, real-time state estimation, and uncertainty handling. These methods have proven effective in managing the relatively predictable dynamics of conventional power systems. However, with the high system dynamics of modern power systems, these models face significant challenges, particularly in terms of modeling complexity, ensuring stability guarantees, and managing the intensive computational burden.

Additionally, as DERs become more prevalent, the operational landscape of power systems has changed drastically. The increased integration of renewable energy sources (RESs) like solar and wind introduces significant variability and uncertainty, complicating the operation and control of modern power systems [10], [11]. Accurately modeling these systems has become increasingly difficult and, in some cases, nearly impossible, particularly for large-scale power and energy networks. As a result, the need for more adaptive and robust control strategies that can handle the inherent uncertainty, and dynamics of modern power systems has become more critical than ever.

*Corresponding authors: V.H. Bui (vhbui@umich.edu) and W. Su (wencong@umich.edu).*

Due to the widespread deployment of wide-area monitoring systems, advanced metering infrastructures, and other monitoring and management technologies, modern power and energy systems have provided vast amounts of data. This data-rich environment paves the way for machine learning (ML)-based optimization approaches to assist in power system operation and control. ML techniques, including neural networks [12] and graph neural networks [13], have been utilized to estimate system behaviors and provide valuable operational insights. However, these model estimations may not always offer reliable solutions, especially given the high levels of uncertainty and complexity in modern power systems. The challenge lies in ensuring the accuracy of these models and developing robust control strategies for real-world scenarios, where system dynamics can change rapidly and unpredictably.

To address these challenges, deep reinforcement learning (DRL) has shown promise in managing flexible energy resources and demand-side management in modern power systems, especially considering the high penetration of DERs [14], [15]. While DRL algorithms have significantly advanced control schemes [16], [17] by solving complex decision-making and optimization problems, they often fall short in guaranteeing safety—a critical concern in the operation of critical infrastructure like power and energy systems. This limitation has led to the development of safe RL as a potential solution. Safe RL not only performs effectively in dynamic environments with high uncertainty but also ensures that decision-making processes are safe, thereby enhancing the reliability and security of power system operations.

In recent years, there has been a growing body of research focused on utilizing safe RL to enhance the efficiency, reliability, and resiliency of power and energy systems. These studies cover a wide range of applications, from small devices and components, such as power converters, smart home systems, and EV/battery controls, to large-scale systems like microgrids and national power networks. The recent literature on safe RL can be broadly classified into two key areas *(i) component-level applications* and *(ii) system-level applications*, as summarized in Table I. This classification highlights the versatility of safe RL in addressing the unique challenges faced at different scales of power system operation.

This paper provides a comprehensive review of the latest research efforts focused on safe RL in power system operation and control, with a particular emphasis on the unique safety requirements of these systems. We delve into safe RL-based approaches across various component-level applications, including grid-connected power converters, EV/battery controls, and smart home and building energy management. Additionally, we explore system-level applications, such as the operation and control of microgrids, power distribution networks, and large-scale power systems. For each approach and application, we discuss the associated bottlenecks, research challenges, and potential opportunities in the operation and control of power systems. We also outline future directions aimed at enhancing the performance of safe RL algorithms and improving system efficiency and reliability. The major

TABLE I
SAFE RL APPLICATIONS IN POWER AND ENERGY SYSTEMS

| Application level | Application area | Focus | References |
|---|---|---|---|
| Component level control | Grid-connected power converter | Voltage control and parameters' optimization | [18], [19], [20] |
| | | Decentralized controller | [21], [22], [23], [24] |
| | | Cooperative control with multiagent systems | [25], [26], [27] |
| | EVs and battery | Charging/discharging schedule | [28], [29], [30] |
| | | Optimal operating range | [31], [32], [33], [34] |
| | | Online optimization | [35], [36], [37] |
| | Smart home and building | Operation cost minimization | [38], [39], [40] |
| | | HVAC control | [41], [42], [43] |
| | | Maintaining customer comfort level | [44] |
| System level control | Microgrids | Islanded microgrid operation and control | [45], [46], [47] |
| | | Grid-connected microgrid operation and control | [48], [49], [50], [51], [52] |
| | | Networked microgrid operation and control | [53], [54], [55], [56], [57] |
| | Power distribution networks | Volt-VAR control | [58], [59], [60], [61], [62], [63], [64], [65] |
| | | Economic dispatch | [66], [67], [68], [69] |
| | | Resiliency and load restoration | [70], [71], [72], [73] |
| | Power generation and transmission systems | Coordinated frequency regulation | [74], [75], [76], [77] |
| | | Stability guarantees | [78], [79], [80], [81] |
| | | Dispatch strategy | [82], [83], [84], [85], [86] |

contributions of this study are as follows:
- Providing an overview of current trends and advancements in safe RL for power and energy systems.
- Reviewing a substantial number of recent studies in the domain of safe RL applications to provide a comprehensive understanding of the state-of-the-art.
- Identifying the limitations and opportunities of existing safe RL approaches, highlighting areas where further research is needed.
- Identifying key challenges and suggesting future directions for improving safe RL algorithms and their applications in power and energy systems, aiming to enhance their effectiveness and reliability in real-world scenarios.

The remainder of this paper is organized as follows: Section 2 provides a background on RL, safe RL, and common open-source tools used for developing safe RL applications. Section 3 discusses the application of safe RL at the component level, focusing on power converters, battery management, and home/building energy management systems. Section 4 presents system-level applications, exploring the use of safe RL in microgrids and large power systems. Section 5 presents a discussion of the challenges faced in implementing safe RL and outlines future directions for research and development. Finally, Section 6 concludes the paper, summarizing the key insights of the study.

## II. Background of Reinforcement Learning and Safe Reinforcement Learning

In this section, we summarize the background of various RL algorithms, highlighting their limitations in terms of safety guarantees. These limitations underscore the critical need for advanced RL algorithms that can ensure the safety of the learning agents' actions, which has led to a recent surge in research focused on safe RL. We provide a background and summary of safe RL algorithms, detailing how they address the safety concerns inherent in conventional RL approaches. Additionally, we recommend several open-source platforms that are well-suited for implementing safe RL, offering researchers and practitioners practical tools to develop and test these advanced algorithms.

### A. Reinforcement Learnings

RL and DRL are critical research areas within the machine learning domain, where an agent learns to make decisions by executing a series of actions in a dynamic environment to maximize cumulative rewards. The basic components of RL include an agent and an environment, as illustrated in Fig. 1. Unlike conventional approaches, RL does not require a predefined model of environmental dynamics; instead, it operates based on the concepts of states, actions, policies, and rewards [87]-[90]. The agent observes the current state of the environment and selects the most appropriate action using an optimal policy. The environment then processes the agent's action, updates to a new state, and calculates a corresponding reward. This reward is subsequently fed back into the learning agent, allowing it to refine and optimize its policy for future actions.

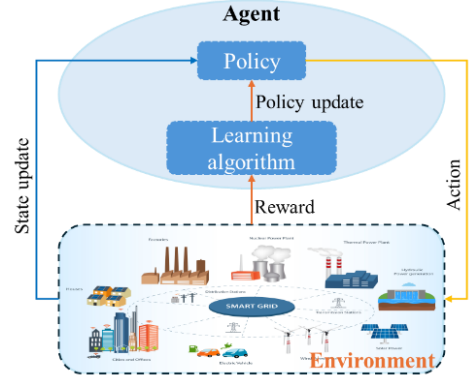

Fig. 1. Basic operation framework of RL agents.

There are three major approaches to designing RL or DRL algorithms, including model-based, policy-based, and value-based methods [88]-[90], as summarized in Fig. 2.
- Model-based approach: This approach involves building a model of the environment that can predict the next state and the reward for each action taken from a given state. The model is then used to plan and decide the best actions to take. This approach is suitable for scenarios where interactions with the environment are costly or risky.
- Policy-based approach: This approach focuses directly on learning the policy that maps states to actions. The policy can be deterministic, where a state directly maps to an action, or stochastic, where a state maps to a probability distribution over actions. This approach is suitable for high-dimensional or continuous action spaces.
- Value-based approach: This approach learns the value of being in a given state, or taking an action in a state, without explicitly learning the policy. The value function estimates how beneficial it is to be in a certain state or to perform a certain action in a state, considering future rewards. This approach is suitable for problems with discrete or finite action spaces.

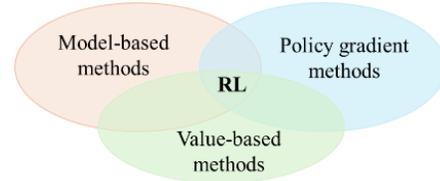

Fig. 2. Three major model formulations for RL agents.

Mathematically, decision making problem often modeled as a Markov decision process (MDP), which is defined by a combination of a state space $\mathcal{S}$, an action space $\mathcal{A}$, the transition probability function $\mathbb{P}(.|s,a): \mathcal{S} x \mathcal{A} \rightarrow \Delta(\mathcal{S})$ maps state-action pair $(s,a) \in \mathcal{S} x \mathcal{A}$ to a distribution over a state space, and a reward function $r(s,a): \mathcal{S} x \mathcal{A} \rightarrow \mathbb{R}$. It is important to note that the state and action space can be either discrete or continuous, which significantly influences the complexity and methods used for solving the MDP.

The long-term goal of an agent is to determine an optimal policy $\pi^*$ that maximizes the expected infinite horizon discounted rewards or the cumulative rewards $J(\pi)$, as given in (1).

$$\pi^* \in \underset{\pi}{argmax}\, J(\pi) = \mathbb{E}_{s_0 \sim \mu_0} \mathbb{E}_\pi \sum_{t=0}^{\infty} \gamma^t r(s_t, a_t) \quad (1)$$

The first term $\mathbb{E}_{s_0 \sim \mu_0}$ represents the expectation over the initial state $s_0$, which is drawn from the state space according to the distribution $\mu_0$. The second term $\mathbb{E}_\pi$ represents the expectation over the action $a_t$ taken using the policy $\pi(.|s_t)$. The parameter $\gamma \in (0,1)$ is the discount factor, which assigns higher weight to short-term rewards (current rewards) compared to long term rewards (future rewards).

For most policy-based RL algorithms, a policy is defined as $\pi_\theta$, where the parameter $\theta \in \Theta \subseteq \mathbb{R}^K$. With this parameterization, the objective in equation (1) can be rewritten as a function of the policy parameter $\theta$, as shown in (2). The optimal policy can be obtained by determining the optimal $\theta^*$, which maximize the objective function.

$$\theta^* \in \underset{\theta \in \Theta}{argmax}\, J(\theta) \quad (2)$$

To find the solution of (2), gradient ascent method is widely used in recent studies, as given in (3).

$$\theta \leftarrow \theta + \eta \nabla J(\theta) \quad (3)$$

where $\eta$ is the step size and the gradient $\nabla J(\theta)$ is computed using policy gradient theorem, as shown in (4).

$$\nabla J(\theta) = \sum_{s \in S} \mu_\theta(s) \sum_{a \in \mathcal{A}} \pi_\theta(a|s) Q_{\pi_\theta}(s,a) \nabla_\theta \ln \pi_\theta(a|s) \quad (4)$$

where $\mu_\theta(s) \in \to \Delta(S)$ is the on-policy state distribution, denoting the fraction of time steps spent in each state $s \in S$. The stochastic policy is presented as $a \sim \pi_\theta(.|s)$.

In RL, various algorithms have been introduced and have gained prominence due to their effectiveness and widespread application across different domains. These include Q-learning, dyna Q-learning, deep Q network (DQN), and deep deterministic policy gradients (DDPG) that combines the benefits of both policy-based and value-based methods.

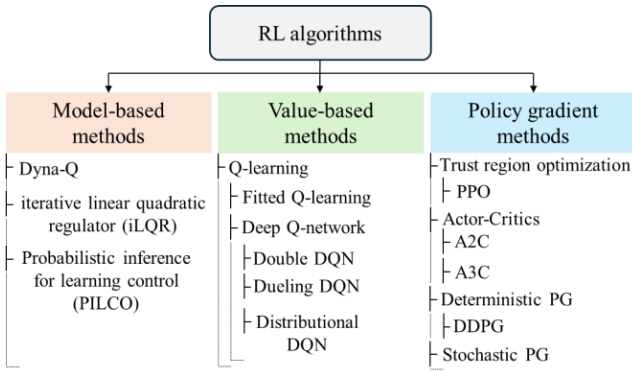

Fig. 3. Methods and algorithms of RL.

Additionally, A2C, A3C, SAC have also been widely adopted for their robustness and efficiency in continuous action spaces. Most popular algorithms of conventional RL are illustrated in Fig. 3.

### B. Safe Reinforcement Learning

Despite the tremendous success of RL/DRL algorithms in various environments, applying these algorithms to real-world applications in critical infrastructure, such as power and energy systems, still faces significant challenges. One of the major concerns is ensuring operational safety or satisfying constraints during the decision-making process. State-wise constraints are among the most common and challenging constraints in safe RL, as they require the agent to avoid unsafe states throughout the learning and operational phases. Ensuring compliance with state-wise constraints is crucial for many applications, including power system operation and control, EV/battery management, and system frequency regulation. In this section, we provide a brief introduction to safe RL and discuss popular safe RL algorithms that have been developed and studied in the literature [91]-[94].

The standard goal for MDP, is to learn a policy $\pi$ that maximizes the cumulative reward, $J(\pi)$, as expressed in (1). In safe RL, a constrained Markov decision process (CMDP) is introduced as an MDP augmented with constraints that restrict the set of allowable policies for the MDP. In other words, CMDP introduces a set of cost functions, $C_1, C_2, \ldots, C_m$, where $C_i: S \times \mathcal{A} \times S \to \mathbb{R}$ maps the state action transition tuple into a cost value. Similar to (1), we denote $J_{C_i}(\pi) = \mathbb{E}_{\pi \sim \tau}\left[\sum_{t=0}^{\infty} \gamma^t C_i(s_t, a_t, s_{t+1})\right]$ as the cost measure for policy $\pi$ with respect to cost function $C_i$. Therefore, the set of feasible stationary policies for CMDP is then defined as follows.

$$\Pi_C = \{\pi \in \Pi | \forall i, J_{C_i}(\pi) \leq d_i\} \text{ with } d_i \in \mathbb{R} \quad (5)$$

In CMDP, the objective is to select a feasible stationary policy $\pi_\theta$ that maximizes the performance measure:

$$\underset{\theta}{max}\, J(\pi_\theta)\ with\ \pi_\theta \in \Pi_C \quad (6)$$

In safe RL, we are specifically interested in a special type of CMDP where the safety specification is to persistently satisfy a hard cost constraint at every step, which is called state-wise constrained Markov decision process (SCMDP). Similar to CMDP, SCMDP also uses the set of cost functions, $C_1, C_2, \ldots, C_n$, to evaluate the instantaneous cost for state action transition tuple. SCMDP requires the cost for every state action transition satisfies a hard constraint. Hence, the set of feasible stationary policies for SCMDP is defined in (7).

$$\bar{\Pi}_C = \{\pi \in \Pi | \forall (s_t, a_t, s_{t+1}) \sim \tau, \forall i, C_i(s_t, a_t, s_{t+1}) \leq \omega_i\} \text{ Where } \tau \sim \pi \text{ and } \omega_i \sim \mathbb{R} \quad (7)$$

The objective for SCMDP is to find a feasible stationary policy from $\bar{\Pi}_C$ that maximizes the performance measure, as given in (8).

$$\max_{\theta} J(\pi_\theta) =, with\ \pi_\theta \in \overline{\Pi}_C \qquad (8)$$

Major safe RL methods and algorithms are summarized into five major classes, as shown in Fig. 4. Among these, the five most prominent algorithms are utilized in smart grid applications, with their pros and cons tabulated in Table II.

TABLE II
BASIC SAFE RL MODELS AND PROS & CONS

| Algorithms | Pros | Cons |
|---|---|---|
| Constrained Policy Optimization (CPO) | - Theoretically sound with guarantees of monotonic improvement. | - Computationally intensive due to solving constrained optimization problems.<br>- Requires careful hyperparameter tuning. |
| Penalty Parameterized Policy Optimization (P3O) | - Balances reward maximization and constraint satisfaction dynamically<br>- More computationally efficient than CPO. | - May require careful tuning of penalty parameters.<br>- Can oscillate if penalties are not properly adjusted. |
| Proximal Policy Optimization with Lagrangian (PPOLag) | - Simpler and more stable than TRPO-based methods.<br>- Effective in practice with good empirical performance. | - Balancing reward and constraints can be challenging.<br>- Requires careful tuning of Lagrange multipliers. |
| Trust Region Policy Optimization with Lagrangian (TRPOLag) | - Combines the stability and performance guarantees of TRPO with constraint satisfaction. | - Computationally intensive.<br>- Requires solving a constrained optimization problem, which can be complex. |
| Soft Actor-Critic with Lagrangian (SACLag) | - Combines the benefits of entropy maximization (exploration) with constraint satisfaction.<br>- Effective in continuous action spaces. | - Computationally demanding.<br>- Requires tuning of both SAC-specific parameters and Lagrange multipliers. |

## C. Common Safe RL Frameworks and Implementation

With the growing emphasis on safety concerns within both the industry and academic communities, there has been a rapid development of open-source safe RL frameworks [95]-[99]. The key features of these frameworks are summarized in Fig. 5.

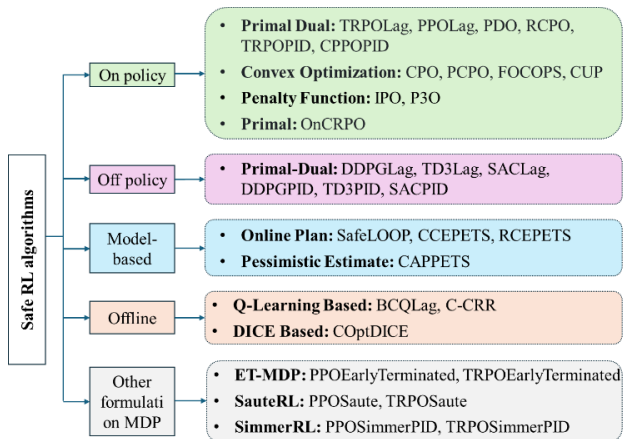

Fig. 4. Main safe RL algorithms and their relationships [95].

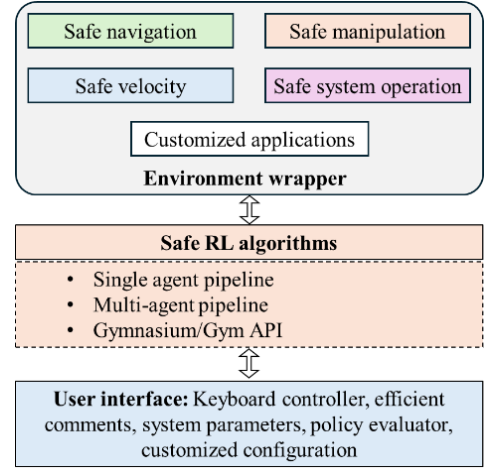

Fig. 5. Key feature available of safe RL frameworks.

They are commonly applied in areas such as safe navigation (e.g., car driving), safe manipulation (industrial applications), safe velocity control (e.g., humanoids), and safe system operation (industrial applications). These frameworks also offer users the flexibility to develop their own applications and environments, enabling easy implementation and customization to meet specific needs.

These frameworks typically provide a wide range of safe RL algorithms, allowing users to freely choose the most suitable algorithm for a given environment. They also often facilitate user implementation with features such as user interfaces/inputs, including keyboard controllers, user comments, policy evaluators, and more. In this study, we have listed five safe RL development frameworks available online as follows.

i. *OmniSafe [95]* is an infrastructural framework designed to accelerate safe RL research. It provides a comprehensive and reliable benchmark for safe RL algorithms, along with an out-of-the-box modular toolkit for researchers. OmniSafe stands as the inaugural unified learning framework in the realm of safe RL, aiming to foster the growth of the safe RL learning community.

ii. *Safe RL Baselines [96]* is a GitHub repository offering various safe RL baselines and benchmarks, including both single-agent and multi-agent RL, specifically for safe RL research. The repository introduces dozens of algorithms with detailed implementations.

iii. *Beaver [97]* is a highly modular open-source safe RL from human feedback (RLHF) framework developed by the PKU-Alignment team at Peking University. It aims to provide training data and a reproducible code pipeline for alignment research, particularly for constrained alignment in large language models (LLM) research via safe RLHF methods.

iv. *VSRL-Framework [98]* is the verifiably safe RL framework. The concept of verifiable safety means that these safety constraints are supported by formal, computer-checked proofs.

v. *SafeDreamer [99]* focuses on safe RL with world models, incorporating Lagrangian-based methods into

world model planning processes within the superior Dreamer framework.

Table III provides a comprehensive comparison of the key features of the five frameworks. It can be seen that framework has its own pros and cons. However, OmniSafe stands out for its flexibility and is favored by many researchers in recent studies.

TABLE III
FRAMEWORK/REPOSITORY KEY FEATURE COMPARISON

| Framework | Multiple agents | Various algorithms | Highly modular framework | Parallel computing acceleration | Customized environments |
|---|---|---|---|---|---|
| OmniSafe [95] | √ | √ | √ | √ | √ |
| Safe RL baselines [96] | √ | √ | x | x | √ |
| Beaver [97] | √ | √ | √ | x | √ |
| VSRL-framework [98] | √ | √ | x | x | x |
| SafeDreamer [99] | √ | √ | x | x | x |

## III. SAFE RL IN COMPONENT & RESIDENTIAL LEVEL APPLICATIONS

Safe RL is an advanced approach that integrates safety constraints into the RL paradigm, ensuring that autonomous systems operate not only efficiently but also within predefined safety limits. This approach is particularly crucial in applications where the consequences of unsafe actions can be severe, such as in power and energy systems, electric vehicles (EVs), and smart home and building environments. In this section, we explore recent studies on component and residential-level applications of safe RL, focusing on its utilization in power converters, EVs and batteries, and smart homes/buildings. These applications, which are critical to the safe and reliable operation of modern power systems, are summarized in Fig. 6.

### A. Grid-Connected Power Converters

With the significant increase in the penetration of DERs, grid-connected converters have become essential for interfacing between various types of DERs and the electrical grid. Their primary function is to convert the electrical power generated by these DERs into a form compatible with the grid's requirements. This conversion typically involves adjusting the voltage, current, and frequency of the power to match those of the grid.

Safe RL has been widely employed to enhance voltage control and quality [18], [20], [21], ensuring that power is delivered reliably and within safe voltage ranges. Conventional control methods may struggle with the dynamic and uncertain nature of modern power grids. Safe RL addresses these challenges by continuously learning and adapting to changing conditions, optimizing controller parameters to maintain stability and efficiency. A safe RL-based fast time-scale voltage control is proposed in [18] by quickly adjusting reactive power. The authors in [20] have proposed a learning-based finite control set model predictive control (FCS-MPC) to optimize the performance of DC-DC buck converters in DC microgrids. This method addresses the challenge of optimally designing weighting coefficients in the FCS-MPC objective function, improving overall voltage regulation and efficiency. However, high penetration of power-electronic interfaces also requires complex control laws, making decentralized control a viable but challenging approach. A promising method is the use of safe RL to train neural network controllers, ensuring system stability through specific design constraints [19].

Moreover, decentralized control methods provide an ability to reduce communication burdens and enhance system robustness. The authors in [22] have highlighted the limitations of centralized voltage regulation approaches, which, while optimal, suffer from high communication demands. Instead, decentralized methods using DRL can achieve optimal control with lower communication requirements and guaranteed safety constraints. The authors in [23] have addressed the challenges of battery-based multilevel inverters, which are known for their safety and efficiency benefits. Another relevant study [21] have discussed the control of MGs, which rely heavily on power electronic converters. This study emphasized the need for comprehensive testing of new control concepts and automatic tuning of voltage source inverters operating in standalone, grid-forming modes. Decentralized control approaches are crucial for managing the diverse topologies and operational requirements of MGs.

Additionally, the concept of multi-agent system (MAS) is also introduced to cooperation of multiple decentralized controllers. MASs offer innovative solutions for voltage control in power systems. These methods distribute control tasks across multiple agents, which can operate independently or collaboratively. The authors in [25] have introduced a safe multi-agent deep RL algorithm for real-time control of inverter-based Volt-Var in distribution grids, considering communication delays and minimizing network power loss while maintaining voltage stability. Similarly, the authors in [26] have proposed a multi-agent DRL-based autonomous input voltage sharing control for input series output-parallel dual active bridge converters. This approach tackles the uncertainties of DC MGs, power balance issues, and converter stress minimization, showcasing the versatility of MAS in addressing complex power system challenges. In another study [27], secondary QU droop control is enhanced using multi-agent RL to integrate converter-interfaced generators (CIGs) into electrical grids more effectively. This method increases energy flexibility and maximizes hosting capacity by enabling decentralized control of secondary voltage through converters. Major approaches are summarized in table IV.

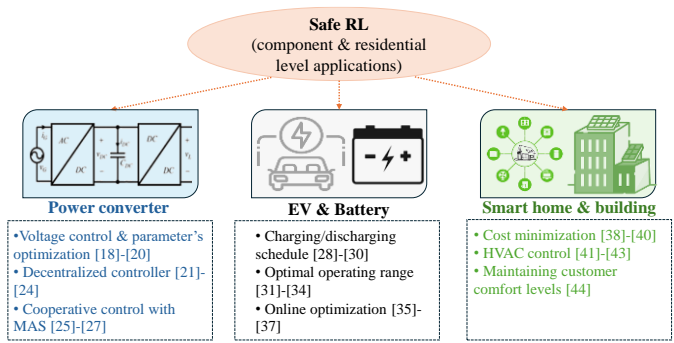

Fig. 6. Safe RL for operation of power converters, EV & battery, and smart home & buildings.

TABLE IV
SAFE RL APPLICATIONS IN POWER CONVERTERS

| Application area | Focus | Contributions |
|---|---|---|
| Voltage control | Fast time-scale voltage control in power systems [18] | Proposed a safe RL-based method for quickly adjusting reactive power, enhancing voltage control and quality |
| Voltage control | Decentralized control for voltage stability [19] | Developed a method using safe RL to train neural network controllers, ensuring system stability through specific design constraints |
| Voltage regulation | Optimization of DC-DC buck converters in DC MGs [20] | Proposed a learning-based FCS-MPC approach to optimally design weighting coefficients, improving voltage regulation and efficiency |
| Decentralized-based voltage regulation | Control of MGs with power electronic converters [21] | Emphasized the need for comprehensive testing and automatic tuning of voltage source inverters in standalone, grid-forming modes |
| Decentralized-based voltage regulation | Decentralized control with low communication requirements [22] | Addressed the limitations of centralized voltage regulation, proposing decentralized DRL methods with lower communication demands and guaranteed safety constraints |
| Decentralized-based voltage regulation | Control of battery-based multilevel inverters [23], [24] | Discussed the challenges and solutions for controlling battery-based multilevel inverters, focusing on safety and efficiency benefits |
| Multi-agent system-based control methods | Real-time Volt-Var control in distribution grids [25] | Introduced a safe multi-agent deep RL algorithm to handle communication delays, minimize power loss, and maintain voltage stability |
| Multi-agent system-based control methods | Input voltage sharing for DC MGs [26] | Proposed a multi-agent DRL-based autonomous control method for dual active bridge converters, addressing power balance and stress minimization |
| Multi-agent system-based control methods | Secondary QU droop control for CIGs integration [27] | Enhanced secondary voltage control using multi-agent RL, increasing energy flexibility and maximizing hosting capacity |

### B. Electrical Vehicle and Battery Operation

This section focuses on the integration of safe RL for optimizing battery usage and enhancing the efficiency and reliability of electric vehicle systems. Safe RL plays a vital role in optimizing energy usage and extending battery life. Battery management systems must balance the demands of charging and discharging cycles to prevent degradation and ensure safety. Safe RL can dynamically adjust these cycles based on real-time data, enhancing the longevity of the battery and the efficiency of the vehicle. This is particularly important as EVs become more prevalent and the demand for reliable, long-lasting batteries increases. Safe RL ensures that EVs can operate under various conditions without compromising safety or performance.

Various studies have discussed charging/discharging control for safety operation of EVs or energy storage systems (ESSs) [28]-[30]. Efficient management of EV charging and discharging is essential to optimize the use of energy resources while ensuring the safety and readiness of EVs. A safe RL-based charging scheduling strategy for a residential microgrid system has been proposed in [29], considering different types of EVs and vehicle-to-grid (V2G) modes. The problem is formulated as a constrained Markov decision process (CMDP) and solved using a constrained soft actor-critic algorithm, with a safety filter ensuring safe operations. The authors in [28] have developed a model-free approach based on safe deep RL for EV charging/discharging scheduling to minimize costs while guaranteeing full charge upon departure. This approach does not require domain knowledge and directly learns the optimal schedules using a deep neural network (DNN). The authors in [30] have presented a safe RL framework that embeds rule-based local and global shields to supervise agent actions in a large-scale partially-observable CMDP for EV charging control. This framework ensures safety during training and execution while finding near-optimal policies.

Another aspect of operation of EVs/battery is to maintain the optimal operating range of energy storage systems, which is critical to enhance efficiency and prolong the lifespan of batteries. Safe RL has been employed to develop strategies that manage energy systems effectively within their safe operational limits. Battery energy management system has been developed by [31], [32]. In [31] the author has introduced a penalty term to guide the agent to learn how to operate the battery charging and discharging in a safety boundary, 10-90%, a high penalty to apply to avoid making bad action using Q-learning. The authors in [32] also improve the learning agent with DDQN-based management system, which not only leverages all advantages from [31] but also considers system uncertainty. The model could perform well even with system uncertainty.

Focusing more on the safety aspect, the authors in [33] developed an energy management strategy based on deep RL for a hybrid battery system in EVs, aiming to minimize energy loss and enhance safety. A novel reward term is designed to explore the optimal operating range without imposing rigid constraints. The authors in [34] proposed a PASACLag algorithm, a safe hybrid-action RL approach for HEV energy management, addresses both continuous and discrete actions while maintaining battery state of charge within safe limits. It uses a Lagrangian method to separate control objectives and constraints, enhancing safety and performance.

Finally, real-time decision-making and online optimization are vital for the dynamic management of energy systems and EVs. Safe RL techniques have been developed to address the uncertainties and complexities of real-time operations. The authors in [35] have developed a novel DRL framework for online optimization in the energy management of a power-split HEV. The framework combines a DNN-based multi-constraints optimal strategy and a rule-based restraints system to optimize fuel economy and avoid irrational control signals.

TABLE V
SAFE RL APPLICATIONS IN BATTERY & EVS

| Application area | Focus | Contributions |
|---|---|---|
| Charging/discharging control | Safe operation of EVs and ESSs [28] | Developed a model-free safe deep RL approach for EV charging/discharging scheduling to minimize costs while ensuring full charge upon departure without domain knowledge |
| Charging/discharging control | Safe RL-based charging scheduling strategy for residential microgrids [29] | Proposed a CMDP-based safe RL approach for charging scheduling, utilizing a constrained soft actor-critic algorithm and a safety filter for safe operations in various V2G modes |
| Charging/discharging control | Large-scale EV charging control in partially observable environments [30] | Presented a safe RL framework with rule-based local and global shields to supervise agent actions, ensuring safety during training and execution |
| Optimal operating range | Battery management within safe operational boundary [31] | Introduced a penalty term in Q-learning to guide agents in safe battery charging/discharging within 10-90% boundaries to avoid making harmful actions |
| Optimal operating range | DDQN-based battery management system under uncertainty [32] | Improved upon previous models by incorporating system uncertainty, maintaining safe operations within specified limits using a DDQN-based management system |
| Optimal operating range | Hybrid battery system energy management in EVs [33] | Developed a deep RL strategy for managing a hybrid battery system, designing a novel reward term to explore optimal operating ranges without rigid constraints |
| Optimal operating range | Safe hybrid-action RL for HEV energy management [34] | Proposed PASACLag algorithm, a hybrid-action RL approach using a Lagrangian method to maintain battery state of charge within safe limits, addressing both continuous and discrete actions |
| Online optimization | Online optimization of energy management in power-split HEVs [35] | Developed a DRL framework for real-time optimization combining a DNN-based strategy and a rule-based system to enhance fuel economy and prevent irrational control signals |
| Online optimization | Real-time EV charging management in charging stations [36] | Proposed a DRL-based allocation approach for real-time EV charging, integrating safety modules to ensure secure and accurate allocation amidst grid and EV-related uncertainties |
| Online optimization | Dynamic dispatch of battery energy storage systems in microgrids using RL and Monte-Carlo tree search [37] | Introduced an RL solution with Monte-Carlo tree search and domain knowledge to enhance computational efficiency and incorporate lifecycle degradation costs in the optimization model |

The authors in [36] have proposed a DRL-based allocation approach for real-time management of EV charging in a charging station, optimizing EVs' charging while addressing grid and EV-related uncertainties. Safety modules are integrated to ensure security and allocation accuracy. The authors in [37] proposed an RL solution augmented with Monte-Carlo tree search and domain knowledge for dynamic dispatch of battery energy storage systems in microgrids. This approach improves computation efficiency and incorporates lifecycle degradation costs into the optimization model. Major research problems are summarized in table V.

*C. Residential Smart Homes and Buildings*

In smart homes and buildings, safe RL is applied to optimize energy consumption while maintaining occupant comfort. For instance, by integrating safe RL into heating, ventilation, and air conditioning (HVAC) systems, these buildings can autonomously adjust temperature and airflow to minimize energy costs without sacrificing comfort. Furthermore, safe RL helps manage the complex interplay of various smart devices and systems within a building, ensuring that energy usage is optimized holistically. This not only reduces operational costs but also contributes to sustainable living by lowering the overall energy footprint of residential and commercial buildings.

These applications of safe RL can be grouped into three main categories, including cost minimization, HVAC control, and ensuring customer comfort level. Each category addresses unique challenges and leverages safe RL to enhance efficiency, reduce costs, and improve user experience in home and building energy systems.

Cost minimization is a critical objective in smart home and building energy management, aiming to reduce electricity expenses while maintaining efficient operation of various energy systems. The authors in [38] have proposed a data-driven approach for multi-energy management of smart homes, formulating the problem as a cost minimization task with hard constraints. Utilizes a safe RL approach with primal-dual optimization policy search and a DNN for electricity price forecasting. Demonstrates superior performance in minimizing energy costs and satisfying constraints compared to existing methods. The authors in [39] have addressed the real-time optimization of EV charge/discharge scheduling in smart home applications to minimize electricity costs using proximal policy optimization considering the system uncertainty. The authors in [40] have proposed a multiagent RL approach for residential multicarrier energy management using Q-learning and scenario-based methods to handle uncertainties. This model aims to minimize the operation cost for consumers compared to conventional optimization-based programs.

Another aspect of smart home or building control focuses on efficient control of HVAC systems, which is essential for maintaining comfortable indoor environments while optimizing energy use. The authors in [41], [43] have developed safe building HVAC control via batch RL, proposing guided exploration and conservative Q-learning to ensure safety and optimize HVAC policies. The algorithm shows significant reductions in ramping and daily peak compared with the conventional rule-based methods. The authors in [42] have proposed a multi-agent DRL framework for building energy systems, using dueling double deep Q-network and value-decomposition network for cooperative optimization. This model utilizes the prioritized experience replay to accelerate convergence and maintain stability, achieving multi-objective optimization and significant improvements in renewable energy usage and cost reduction.

TABLE VI
SAFE RL APPLICATIONS IN SMART HOMES AND BUILDINGS

| Application area | Focus | Contributions |
|---|---|---|
| Cost minimization | Multi-energy management in smart homes [38] | Proposed a data-driven approach using safe RL with primal-dual optimization policy search and a DNN for electricity price forecasting, achieving superior performance in minimizing energy costs and satisfying constraints |
| Cost minimization | Real-time optimization of EV charge/discharge scheduling in smart homes [39] | Developed a safe RL-based method using proximal policy optimization to minimize electricity costs in smart home applications while considering system uncertainty |
| Cost minimization | Residential multicarrier energy management [40] | Proposed a multiagent RL approach using Q-learning and scenario-based methods to handle uncertainties, aiming to minimize operation costs for consumers compared to conventional optimization-based programs |
| HVAC control | Safe HVAC control in buildings [41], [43] | Developed a batch RL algorithm with guided exploration and conservative Q-learning for safe HVAC control, achieving significant reductions in ramping and daily peak compared with conventional rule-based methods |
| HVAC control | Cooperative optimization of building energy systems [42] | Proposed a multi-agent DRL framework using dueling double deep Q-network and value-decomposition network for building energy systems, enhancing renewable energy usage and cost reduction while maintaining stability and convergence |
| Customer comfort level | Resilient proactive scheduling in commercial buildings during extreme weather events [44] | Introduced a safe RL-based strategy combining deep-Q-network and conditional-value-at-risk methods to optimize control decisions, maximizing customer comfort while minimizing energy reserve costs |

Beside the cost reduction, the customer comfort level is also an important aspect of smart home and building energy management, especially during extreme events. The authors in [44] have proposed a safe RL-based resilient proactive scheduling strategy for commercial buildings during extreme weather events. This approach combines deep-Q-network and conditional-value-at-risk methods to handle uncertainties, optimizing control decisions to maximize customer comfort while minimizing energy reserve costs. Major research problems are summarized in table VI.

## IV. SAFE RL FOR SYSTEM LEVEL APPLICATIONS

Safe RL has demonstrated significant importance in the operation of devices and components, electrification applications, and home/building energy management within the power and energy system. In this section, we explore the application of safe RL at the system level operation and control, including MGs, distribution networks, and large power systems. Each of these domains highlights specific areas where safe RL can be effectively utilized to enhance operational efficiency, ensure safety, and address the unique challenges inherent to each system, as summarized in Fig. 7.

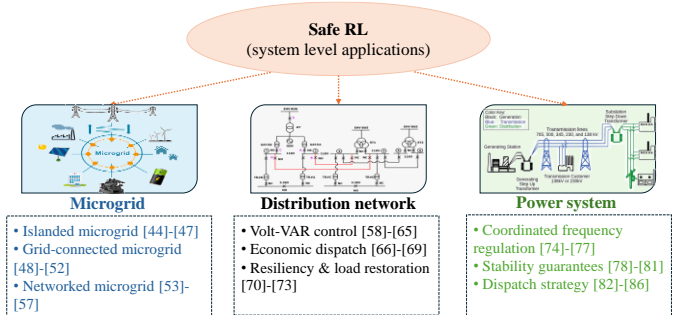

Fig. 7. Safe RL for operation of microgrid, distribution network, and large power system.

### A. Microgrid Operation and Control

MGs, which are comprised of DERs, renewable energy sources, EVs, and various kinds of loads, have emerged as a promising solution for enhancing energy reliability and sustainability. These microgrids can operate in both grid-connected and islanded modes, allowing them to either interact with the main power grid or function independently. However, the high penetration of DERs and renewable energy sources introduces significant challenges in MG planning and operations, particularly due to the inherent uncertainties associated with renewable energy generation and load demands. These challenges are especially pronounced in islanded mode, where the MG must maintain stability and reliability without the support of the main grid.

To address these complexities, safe RL has been increasingly applied to MG operations [100]-[106]. Safe RL provides robust strategies for managing uncertainties and ensuring that MGs operate efficiently and safely under varying conditions. Studies on safe RL-based operation and control of MG systems can be grouped into three different classes (i) islanded operation mode, (ii) grid-connected operation mode, and (iii) interconnected multi-MG.

Safe RL offers a powerful solution for managing MGs in islanded operation mode, ensuring that MGs can dynamically adapt to fluctuations in energy supply and demand while maintaining stability and reliability. The authors in [45] have presented a novel two-stage learning framework for scheduling and dispatching DERs in an islanded MG during utility grid outages. The authors in [46] have developed an online energy management approach using safe DRL to address the challenges of intermittent renewable energy in microgrids. The method formulates the problem as a constrained Markov decision process (CMDP) to ensure safety and minimize operational costs. The authors in [47] have addressed the challenges of using DRL for microgrid energy management by formalizing the problem as a CMDP. The proposed interior-

10TABLE VII
SAFE RL APPLICATIONS IN MG OPERATIONS

| Application area | Focus | Contributions |
| --- | --- | --- |
| Islanded MG operation | Scheduling and dispatching DERs in an islanded MG during grid outages [45] | Presented a two-stage learning framework for scheduling and dispatching DERs, ensuring reliable operation during utility grid outages. |
| Islanded MG operation | Energy management in microgrids with intermittent renewable energy [46] | Developed an online energy management approach using safe DRL formulated as a CMDP, ensuring safety and minimizing operational costs in the presence of intermittent renewable energy. |
| Islanded MG operation | DRL for microgrid energy management [47] | Addressed the challenges of using DRL by formalizing the problem as a CMDP and proposed an interior-point policy optimization method to enhance learning, ensuring constraint satisfaction and cost-effectiveness. |
| Grid-connected MG operation | Real-time operation of MGs to minimize operation losses [48] | Developed a new RL model with a DNN-based surrogate model for accurate loss estimation and system dynamics, optimizing setpoints for various DGs within predefined safe operation values. |
| Grid-connected MG operation | Multi-energy management systems in grid-connected MGs [49], [50] | Introduced safe RL-based models focusing on increasing initial utility and improving constraint accuracy for multi-energy management systems. |
| Grid-connected MG operation | Optimization of energy flow in integrated power and gas systems [51] | Proposed a model-free safe DRL approach using a constrained soft actor-critic algorithm to optimize energy flow, ensuring real-time operation feasibility and safety. |
| Grid-connected MG operation | Safe RL methods for multi-energy management [52] | Proposed two novel online model-free safe RL methods, SafeFallback and GiveSafe, for hard-constraint satisfaction guarantees during training and deployment in multi-energy management systems. |
| Multi-microgrid operation | Internal trading in networked microgrids [53] | Developed a SAC-based optimization model for internal trading, setting safe internal trading prices to minimize load shedding and encourage agent participation. |
| Multi-microgrid operation | Decentralized economic frequency control for isolated networked MG systems [54] | Proposed a decentralized economic frequency control method using a multi-agent deep RL framework, addressing isolated networked MG systems. |
| Multi-microgrid operation | Optimal power management in networked microgrids [55] | Presented a supervised multi-agent safe policy learning method considering AC power flow and operational constraints, ensuring safe decisions while maintaining privacy and data ownership. |
| Multi-microgrid operation | Peer-to-peer energy trading and internal energy conversion in interconnected MGs [56] | Investigated peer-to-peer energy trading and internal energy conversion using a multi-agent DRL approach in interconnected multi-energy MGs. |
| Multi-microgrid operation | Bi-level optimization in networked MGs [57] | Addressed a bi-level optimization problem where the distribution system operator optimizes energy prices while microgrids minimize costs using a RL approach, preserving MG privacy and solving the non-linear optimization effectively. |

point policy optimization method enhances learning in complex environments, ensuring constraint satisfaction and improved cost-effectiveness.

In the grid-connected operation mode of MG, safe RL is leveraged to optimize control strategies that primarily focus on minimizing the operation cost of the entire system. The authors in [48] have developed a new RL model for real-time operation of MG to minimize the operation losses and cost saving for entire system. The accurate DNN-based surrogate model also develops to provide accurate losses with system dynamic. This study has provided the agent with capability of determine setpoint of various DG in safe operation between minimum and maximum predefined values by operation and safety guide. The authors in [49], [50] have introduced safe RL-based models in multi-energy management systems, focusing on increasing initial utility and improving constraint accuracy. The authors in [51] have proposed a model-free safe DRL approach to optimize energy flow in integrated power and gas systems, addressing the challenges posed by increasing renewable energy integration. The constrained soft actor-critic algorithm ensures real-time operation feasibility with high computational efficiency and safety guarantees. The authors in [52] have proposed two novel online model-free safe RL methods, SafeFallback and GiveSafe, for multi-energy management systems, offering hard-constraint satisfaction guarantees during training and deployment.

Safe RL also offers optimal operation strategies for networked MGs, where multiple interconnected MGs work together to enhance overall system efficiency and resilience. The authors in [53] have developed a SAC-based optimization model for the internal trading problem, aiming to provide internal trading prices that minimize the load shedding amount across the entire system. Safe actions are taken to set appropriate internal trading prices, encouraging more MG agents to participated in the internal trading market. The authors in [54] have proposed a decentralized economic frequency control method for isolated networked MG systems using a multi-agent deep RL framework. The authors in [55] presents a supervised multi-agent safe policy learning method for optimal power management in networked microgrids, considering AC power flow and operational constraints. The method ensures safe and feasible decisions while maintaining privacy and data ownership through distributed consensus-based optimization. The authors in [56] have investigated peer-to-peer energy trading and internal energy conversion in interconnected multi-energy MGs using a multi-agent DRL approach. The authors in [57] have addressed a bi-level optimization problem in networked MGs, where the distribution system operator optimizes energy prices while microgrids minimize their costs. A RL approach preserves MG privacy while solving the non-linear optimization problem effectively. Major research problems are summarized in table VII.

TABLE VIII
SAFE RL APPLICATIONS IN POWER DISTRIBUTION SYSTEMS

| Application area | Focus | Contributions |
|---|---|---|
| Volt-var control | Safe voltage control in distribution grids with high penetration of RESs [62] | Introduced a DNN-assisted projection-based DRL method that embeds a projection algorithm into the training process to ensure both safe and optimal operation. |
| Volt-var control | Safe RL algorithm for VVC in power distribution networks [63] | Proposed a data-driven safe RL algorithm for VVC, introducing a safety layer within the policy neural network and a mutual information regularization technique to enhance constraint satisfaction during training. |
| Volt-var control | Safe off-policy DRL algorithm for VVC problems [64] | Developed a safe off-policy DRL algorithm for VVC, addressing issues with incomplete and inaccurate distribution network models using a constrained soft actor-critic algorithm. |
| Volt-var control | Three-stage inverter-based peak shaving and VVC framework [65] | Proposed a safe DRL-based three-stage inverter framework for peak shaving and VVC, coordinating energy storage systems and photovoltaic systems to reduce peak load, voltage violations, and power loss across different timescales. |
| Economic dispatch | RL controller for economic dispatch in future power systems [66] | Proposed a formally validated RL controller for economic dispatch, adding a safety layer to ensure safe dispatch decisions under high complexity and uncertainty. |
| Economic dispatch | Real-time pricing strategies for distribution network congestion management [67] | Addressed the CMDP problem using a safe DRL framework, learning the environment model and applying adaptive cost constraints for real-time pricing strategies. |
| Economic dispatch | Safe RL algorithm for generation bidding and unit maintenance scheduling [68] | Developed a safe RL algorithm combining RL with a predicted safety filter to handle incomplete information and safety constraints in generation bidding and unit maintenance scheduling in competitive electricity markets. |
| Economic dispatch | Multi-agent RL framework for low-carbon demand management [69] | Proposed a multi-agent RL framework using a consensus multi-agent constrained policy optimization algorithm for low-carbon demand management in distribution networks. |
| Resiliency and load restoration | Grid stability amidst fluctuating power consumption and renewable energy variability [70] | Proposed a novel RL-based method incorporating search-based planning for maintaining grid stability amidst fluctuating power consumption, renewable energy variability, and unpredictable disasters. |
| Resiliency and load restoration | DRL-based planning framework for long-term resilience of power distribution systems [71] | Introduced a DRL-based planning framework to enhance long-term resilience through optimal grid hardening strategies. |
| Resiliency and load restoration | Multi-agent DRL method for optimizing load restoration after power outages [72] | Developed a multi-agent DRL method for optimizing load restoration after power outages, including an innovative invalid action masking technique to manage physical constraints and dimensionality challenges. |
| Resiliency and load restoration | Multi-agent RL approach for load restoration in large-scale power systems [73] | Proposed a multi-agent RL approach addressing the challenge of physical constraint violations in load restoration for large-scale power systems. |

*B. Power Distribution Networks*

Safe RL can extend its application from small-scale MGs to larger systems, such as power distribution systems, where it plays a crucial role in addressing the complexities of modern grid operations. In power distribution systems, safe RL focuses on three critical control aspects (i) volt-var control (VVC), (ii) economic dispatch, and (iii) resiliency and load restoration. By optimizing volt-var control, safe RL ensures voltage stability and efficient reactive power management across the network [58]-[65]. For instance, the authors in [62] have addressed the challenge of safe voltage control in distribution grids with high penetration of inverter-based renewable energy sources (RESs) using DRL. It introduces a DNN-assisted projection-based DRL method that ensures both safe and optimal operation by embedding a projection algorithm into the training process. The authors in [63] have proposed a data-driven safe RL algorithm for VVC in power distribution networks. It introduces a safety layer within the policy NN and a mutual information regularization technique to enhance constraint satisfaction during training. The authors in [64] have introduced a safe off-policy DRL algorithm for VVC problems, addressing issues with incomplete and inaccurate distribution network models using a constrained soft actor-critic algorithm. The authors in [65] have proposed a three-stage inverter-based peak shaving and VVC framework using safe DRL in active distribution systems. It coordinates energy storage systems and photovoltaic systems to reduce peak load, voltage violations, and power loss across different control timescales.

In economic dispatch problems, it enables the cost-effective allocation of generation resources while adhering to safety constraints. The authors in [66] have tackled the challenge of economic dispatch in future power systems by proposing a formally validated RL controller. This model added a safety layer to ensure safe dispatch decisions under high complexity and uncertainty. Optimal power trading can cause various challenges for the economic dispatch problem. Therefore, the authors in [67] have addressed real-time pricing strategies for distribution network congestion management using a safe DRL framework. The proposed algorithm handles the CMDP problem by learning the environment model and applying adaptive cost constraints. Similarly, the authors in [68] have proposed a safe RL algorithm for generation bidding and unit maintenance scheduling in competitive electricity markets. The algorithm combines RL with a predicted safety filter to handle incomplete information and safety constraints. The authors in [69] have proposed a multi-agent RL framework for low-carbon demand management in distribution networks using a consensus multi-agent constrained policy optimization algorithm.

Additionally, safe RL can enhance the resiliency of the distribution system by facilitating rapid load restoration and minimizing downtime during disruptions, ensuring reliable and secure grid operation under varying conditions. The authors in [70] have focused on maintaining grid stability amidst fluctuating power consumption, renewable energy variability, and unpredictable disasters. This study proposed a novel RL-based method incorporating search-based planning for safe power grid management. The authors in [71] have introduced a DRL-based planning framework to enhance the long-term resilience of power distribution systems through optimal grid hardening strategies. Addressing post-outage scenarios, the authors in [72] have proposed a multi-agent DRL method for optimizing load restoration after power outages in distribution systems. It includes an innovative invalid action masking technique to manage physical constraints and dimensionality challenges. Similarly, the authors in [73] have proposed a multi-agent RL approach for load restoration in large-scale power systems, addressing the challenge of physical constraint violations. Major research problems are summarized in table VIII.

### C. Power System Generation and Transmission

Beyond its application in power distribution systems, safe RL has emerged as a promising approach for enhancing the operation of power systems at the generation and transmission levels. This advanced method addresses critical challenges across three key categories, (i) coordinated frequency regulation, (ii) stability guarantees, and (iii) dispatch strategy. In coordinated frequency regulation, safe RL dynamically balances supply and demand, ensuring that system frequency remains within acceptable limits despite fluctuations in generation or load. The authors in [74] have introduced the adaptive and safe-certified DRL (AdapSafe) algorithm for frequency control in power systems with inverter-based RESs. The approach integrates a self-tuning control barrier function and meta-RL to ensure safety and adaptability in dynamic environments. The authors in [75] have proposed a safe RL approach for frequency regulation in power systems, utilizing set-theoretic control techniques to guarantee safety without real-time model predictive control. The authors in [76] have presented a model-free coordinated frequency control framework using safe RL to address the challenges posed by the increasing dynamics and uncertainties in power grids. The load frequency control problem is modeled as a CMDP, enabling sub-second decision-making by an AI agent. The authors in [77] have proposed an RL-based approach for optimal transient frequency control in power systems, ensuring stability and safety using Lyapunov stability theory and safety-critical control. Distributed dynamic budget assignment is introduced to reduce conservatism in controller design, allowing for a broader search space for control policies.

Regarding stability guarantees, safe RL techniques are developed to maintain overall system stability by effectively managing disturbances and preventing cascading failures, which is particularly important in complex, interconnected grid environments. The authors in [78] have introduced a novel safe RL method for emergency load shedding to enhance voltage stability and safety in power systems during severe conditions. The approach utilizes a reward function with a barrier function that penalizes unsafe states, ensuring the system avoids safety bounds. The authors in [79], [80] have proposed a safe RL-based load shedding approach for voltage recovery. The method enhances grid resilience by leveraging real-time data and adapting to unforeseen faults. The authors in [81] have proposed a chance-constrained RL algorithm based on proximal policy optimization for supervisory control in power plants, addressing challenges in enforcing state constraints. The approach uses Lagrangian relaxation to convert the constrained optimization problem into an unconstrained one, ensuring safe and efficient control.

TABLE IX
SAFE RL APPLICATIONS IN POWER SYSTEMS

| Application area | Focus | Contributions |
|---|---|---|
| Coordinated frequency regulation | Frequency control in power systems with inverter-based RESs [74] | Introduced the adaptive and safe-certified DRL (AdapSafe) algorithm, integrating a self-tuning control barrier function and meta-RL for safety and adaptability in dynamic environments. |
| Coordinated frequency regulation | Frequency regulation in power systems using set-theoretic control techniques [75] | Proposed a safe RL approach that guarantees safety without requiring real-time model predictive control, utilizing set-theoretic control techniques. |
| Coordinated frequency regulation | Coordinated frequency control framework for power grids [76] | Developed a model-free safe RL framework for coordinated frequency control, modeling the load frequency control problem as a CMDP for sub-second decision-making by an AI agent. |
| Coordinated frequency regulation | Optimal transient frequency control in power systems [77] | Presented an RL-based approach for transient frequency control using Lyapunov stability theory and safety-critical control, introducing distributed dynamic budget assignment to allow for a broader search space for control policies. |
| Stability guarantees | Emergency load shedding for voltage stability and safety [78] | Introduced a safe RL method for emergency load shedding, utilizing a reward function with a barrier function to penalize unsafe states and ensure system safety. |
| Stability guarantees | Load shedding approach for voltage recovery [79], [80] | Proposed safe RL-based methods for load shedding to enhance grid resilience, using real-time data to adapt to unforeseen faults and enhance voltage recovery. |
| Stability guarantees | Supervisory control in power plants with state constraint enforcement [81] | Developed a chance-constrained RL algorithm based on proximal policy optimization for enforcing state constraints in power plants, using Lagrangian relaxation to convert the constrained problem into an unconstrained one. |
| Dispatch strategy | Economic dispatch for virtual power plants [82] | Presented a model-free economic dispatch approach using an adversarial safe RL framework, focusing on robustness against model inaccuracies and environmental uncertainties. |
| Dispatch strategy | Robustness and security in power system operations [83] | Discussed the vulnerabilities of standard RL algorithms to adversarial attacks in power system operations and the design requirements for data and models in safe RL algorithms to ensure robustness and security. |
| Dispatch strategy | Transmission overload risks in power grids [84] | Addressed transmission overload risks by formulating an online preventive control problem as a CMDP, solved using a safe DRL method enhanced with an edge-conditioned convolutional network and long short-term memory network for improved constraint handling and stability. |
| Dispatch strategy | Active power dispatch in power systems [85], [86] | Proposed a safe RL approach for real-time active power dispatch, focusing on balancing generation and consumption while adhering to safety constraints. |

Finally, in the realm of dispatch strategy, safe RL optimizes the allocation of generation resources by considering both economic efficiency and system reliability, ensuring that power generation and transmission are conducted in the most effective manner while adhering to operational constraints. The authors in [82] have presented a model-free economic dispatch approach for virtual power plants using an adversarial safe RL framework, focusing on robustness to model inaccuracies and environmental uncertainties. The authors in [83] have discussed the vulnerabilities of standard RL algorithms in power system operations, such as emergency control and voltage regulation, when subjected to adversarial attacks. It discusses the design requirements for data and models in safe RL algorithms to ensure robustness and security. The authors in [84] have addresses transmission overload risks in power grids with high renewable energy integration by formulating an online preventive control problem as a CMDP. The problem is solved using a safe DRL method, enhanced by an edge-conditioned convolutional network and long short-term memory network to improve constraint handling and stability. The authors in [85], [86] have proposed a safe RL approach for active power dispatch in power systems, focusing on balancing generation and consumption in real-time while adhering to safety constraints. Major research problems are summarized in table IX.

## V. Discussion and Future Directions

In this section, we delve into two critical aspects of safe RL applications in power and energy systems. First, we address the major challenges that safe RL algorithms currently face. Following this, we explore future directions that could enhance the performance and applicability of safe RL, providing detailed strategies to overcome these challenges.

### A. Challenges in Safe RL Approaches

Conventional safe RL often faces various challenges, mainly stemming from the trade-off between effectively exploring to learn optimal policies and maintaining the agent's safety. These challenges can be categorized into four main groups: (i) safety constraint definition, (ii) computational burden, (iii) scalability, and (iv) explainability.

Safety constraint definition is a significant issue in safe RL. It is often difficult to formally define what constitutes a safe state or action in complex environments, such as power and energy systems. The definition of safety can vary significantly across different operation objectives, or operational modes, with each potentially requiring a unique definition of safety.

The computational burden associated with safe RL is considerable. Ensuring that an RL agent operates within safe boundaries in real-time is computationally demanding, requiring complex modeling and verification processes. These processes can be challenging in high-dimensional or dynamic environments, where real-time safety guarantees and the validation of RL policies within safety constraints are both computationally expensive and sometimes infeasible.

Scalability presents another challenge for safe RL. The need for safe exploration often limits the agent's ability to effectively explore the environment, which can hinder learning speed and efficiency due to increased sample requirements. Furthermore, safe RL methods must ensure robustness to unseen scenarios and adaptability across different environments, a necessity for real-world deployment. This adaptability is especially challenging in multi-agent environments, where ensuring inter-agent safety and balancing cooperation and competition further complicates the learning process.

The lack of explainability in safe RL algorithms is another challenge for real-world applications. While safe RL has demonstrated considerable potential in optimizing and securing power and energy system operations and controls, its decision-making processes are often seen as black boxes. This opacity poses a barrier to the broader adoption of safe RL, particularly in critical infrastructures like power and energy systems, where understanding and trust in the decision-making process are paramount.

These challenges underscore the need for continued research and innovation to develop more robust, efficient, and generalizable solutions in safe RL.

### B. Potential Solutions and Future Directions

To address the challenges faced by conventional safe RL, several innovative solutions have been proposed, each aimed at enhancing the safety, efficiency, and scalability of RL algorithms in power and energy systems.

*Safety constraint definition:* One promising approach is the development of dynamic safety constraints that are adaptive or context-aware. These constraints adjust based on the current operation objective or operation modes of power and energy systems, allowing the RL agent to explore effectively while ensuring it operates safely. Also, incorporating risk measures, such as Conditional Value at Risk (CVaR), into the exploration process is another effective strategy. This technique helps prioritize safer actions when exploring unknown states, reducing the likelihood of unsafe behavior.

*Computational burden:* Leveraging model-based RL allows the agent to learn a model of the environment and simulate outcomes, which minimizes the need for real-time computations while ensuring safety. Additionally, gradient-free optimization techniques can efficiently handle non-linear or non-differentiable safety constraints, further enhancing the agent's ability to maintain safety. Another approach is imitation learning, where the RL agent is pre-trained on safe behaviors demonstrated by a human expert or another trusted agent, reduces the exploration needed during training and enhances safety from the onset. Meta-learning techniques can also be applied, enabling the RL agent to quickly learn safe policies across different tasks or environments.

*Scalability:* For multi-agent systems, centralized controllers can use to monitor and enforce safety constraints across all agents, ensuring safe interactions. Alternatively, decentralized learning algorithms allow agents to coordinate and maintain safety independently while optimizing their objectives, this approaches could help address the issue with scalability issue for safe RL.

*Explainability:* Formal verification techniques offer a robust solution for ensuring safety in RL policies by mathematically proving that certain safety properties hold under all circumstances. Developing methods for synthesizing policies that are guaranteed to be safe by construction can provide additional assurance of safety. Human-in-the-loop RL decision-making, involving either real human intervention or large language models (LLMs) to mimic human reasoning, adds a layer of oversight, ensuring that RL actions align with human values and priorities.

## VI. CONCLUSIONS

Safe RL holds significant potential to enhance the performance of various applications within power and energy systems, offering a robust approach to managing the complexities and uncertainties inherent in modern energy infrastructures. This study has presented a comprehensive review of the latest advancements in safe RL as applied to power system operation and control, highlighting the critical importance of maintaining safety in the increasingly complex and dynamic environment of contemporary power grids. By examining a wide array of safe RL algorithms across diverse applications, from single grid-connected power converters to large-scale power distribution networks, this review underscores the versatility and potential of safe RL in improving the reliability and efficiency of power system operations. The study also identifies key challenges, research bottlenecks, and future opportunities in the field, providing valuable insights for advancing safe RL methodologies. Additionally, this review serves as a foundational resource for ongoing and future research, supporting the development of smarter, safer power systems that are capable of adapting to the evolving demands of the energy landscape.

## ACKNOWLEDGMENT

The author's work was supported by the Michigan Institute for Data& AI in Society (MIDAS), University of Michigan, "Propelling Original Data Science (PODS) Grants."